\documentclass{PoS}

\title{Progress and Issues in Hadronic Theory}

\ShortTitle{Progress and Issues in Hadronic Theory}

\author{\speaker{Eric Swanson}\\
       University of Pittsburgh\\
       E-mail: \email{swansone@pitt.edu}}

\abstract{ A brief review of progress and issues in hadronic theory and phenomenology is presented. New results for the $X(3872)$, $Z_c(3900)$, and $Z_c(4020)$ are discussed and unresolved issues are highlighted. A series of open problems in pQCD, NRQCD, and general phenomenology is given. It is argued that these indicate that the current understanding of hadronic dynamics is poor. In particular old ideas about quark annihilation and factorisation appear to be incorrect, pQCD looks limited in scope, and the convergence of some NRQCD computations appears jeapardised by the relative lightness of the charm quark mass.}

\FullConference{XV International Conference on Hadron Spectroscopy \\
		 4-8/11/2013\\
		 Nara, Japan}

\begin{document}

\section{Introduction}

Hadrons and their interactions continue to challenge our ability to compute and model reliably with quantum chromodynamics (QCD). The wealth of new data and states generated by the B factories over the past decade is now being supplemented by new results from BESIII and the LHCb, along with continuing contributions from Belle.  The new results have clarified some issues, created new ones, and left many others unresolved. Some of these are reviewed here.

\section{Charged Onia}

In examining the properties of the enigmatic $Y(4260)$, the BESII collaboration has seen a new enigma\cite{Zc3900} in the process $e^+e^- \to Y(4260) \to \pi^+\pi^- J/\psi$ in the $\pi^\pm J\psi$ spectrum, called the $Z_c(3900)$. The state was soon confirmed by Belle and CLEO-C\cite{Zc3900}.  Breit-Wigner fits yield
$M = 3899.0 \pm 3.6 \pm 4.9$ MeV with a width $\Gamma = 46 \pm 10 \pm 20$ MeV. The state has generated much interest since, if it is not a threshold or other dynamical effect, it must consist of four quarks. 

\begin{figure} [h]
\includegraphics[width=.5\textwidth]{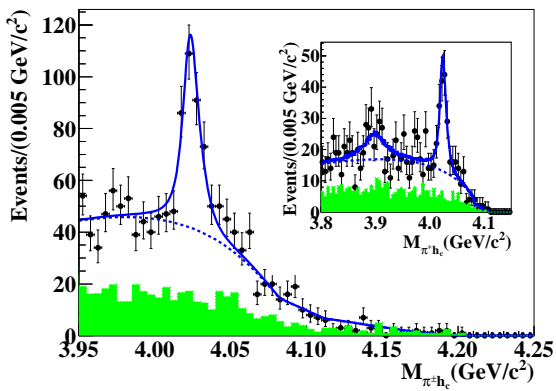} 
\caption{Evidence for the $Z_c(4020)$. Dots with error bars are data; shaded histograms are
normalized sideband background; the solid curves show the total
fit, and the dotted curves the backgrounds from the fit. Figure from Ref. \cite{Zc4020}.}
\label{fig1}
\end{figure}

In the past few months BESIII has found a (possible) partner state in 
$e^+e^- \to \pi^\pm Z_c(4020)^\mp \to \pi^+\pi^- h_c$, called the $Z_c(4020)$\cite{Zc4020} (see Fig. \ref{fig1}). The measured mass is 
$M= 4022.9 \pm 0.8 2.7$ MeV and the width is 
$\Gamma = 7.9 \pm 2.7 \pm 2.6$ MeV. Note that the $Z_c(3900)$ was not seen in this channel. The state is very likely the same as the $Z_c(4025)$ which was seen in 
$D^*\overline{D^*}$ by BESIII\cite{Zc4025}.

These states, along with the older $Z_1(4050)$ and $Z_2(4260)$ (seen in $B \to KZ \to K\pi^\pm\chi_{c1}$) and $Z(4430)$ ($B \to KZ \to K\pi^\pm \psi'$), are a strong indication that four-quark bound states must be taken seriously as possible realisations of QCD dynamics. Furthermore, this story seems to be repeating in the bottomonium sector, where the $Z_b(10610)$ and $Z_b(10650)$ appear as charged bottomonium resonances in $\Upsilon(5S) \to \pi^\pm \pi^\mp \Upsilon(nS)$ and $\pi^\pm \pi^\mp h_b$\cite{Zb}. Recently Belle reports that $J^P = 1^+$ for both states and the preferred decay modes are to $B\overline{B^*}$ and $B^*\overline{B^*}$ respectively\cite{Zb2}. Furthermore, Belle have seen the neutral partner of the $Z_b(10610)$ in $\Upsilon(10860) \to \Upsilon(nS)\pi^0\pi^0$\cite{Zb3}.

Although all of this evidence is encouraging, much work needs to be accomplished before a coherent picture of charged onia emerges. For example, no simple dynamical understanding of these states exists. Are they diquark-diquark, four-quark, or loosely bound? What dynamics provides the binding -- pion or other meson exchange, two-quark interactions, multi-quark interaction, or something new? At the phenomenological level it is disturbing that the $Z_b$ states lie near (and above) $B\overline{B^*}$ and $B^*\overline{B^*}$, while the $Z(4430)$ has not been confirmed by BaBar\cite{babar}, the $Z(4050)$ and $Z(4250)$ look somewhat dubious, the $Z_c(4020)$ is near (and above) $D^*\overline{D^*}$, and the $Z_c(3900)$ is near (and above) $D\overline{D^*}$.

\section{Other New Results}

New developments have not been restricted to spectroscopy. For example, LHCb, which has joined the game, has recently determined that quantum numbers of the $X(3872)$ are $J^{PC} = 1^{++}$\cite{lhcbX}, thereby resolving an old controversy and supporting the $D\overline{D^*}$ molecular picture of this state\cite{ess}. Furthermore, BESIII have measured the process $e^+e^- \to \gamma X(3872)$ and report that, ``the $X(3872)$ might be
from the radiative transition of the $Y(4260)$ rather than from the $\psi(4040)$ or $Y(4360)$, but continuum production ... cannot be ruled out by current data"\cite{YX}. If the decay $Y(4260) \to \gamma X(3872)$ is confirmed it can have important implications for the structure of both states. For example, if the $X$ is a $D\overline{D^*}$ molecule then it is natural to assume that the $Y$ is also a molecule, perhaps a $D\overline{D_1}$\cite{YXtheory}. Alternatively, if the $Y$ is a hybrid state, then this process would require the photon emission to de-excite the gluonic degrees of freedom and create a light quark pair, which seems unlikely. In this case running the process through a $c\overline{c}$ component for the $X$ may be preferred. 

The saga of the pion electromagnetic form factor continues. Recent developments started with BaBar's observation that $Q^2F_\pi(Q^2)$ appears to rise with momentum. This has important implications for the applicability of perturbative QCD to exclusive processes\cite{adam} (including whether the concept of exclusive pQCD even exists). However, Belle have recently repeated the measurement and find a substantially slower rise with momentum\cite{piFF}. Whether one interprets this as the expected pQCD flattening or a rise that disagrees with pQCD appears to be a matter of psychology at present.

LHCb have contributed to  a puzzle concerning the lifetime of the $\Lambda_b$ baryon relative to the $B$ meson. While NRQCD predicts a ratio of nearly unity, old experimental results disagreed with this. The issue now appears resolved in favour of the NRQCD prediction\cite{lhcb}.

The hadronic lattice community is poised to become an important contributor to hadronic phenomenology. The main issues are (i) obtaining light (and chiral) pions, (ii) incorporating continuum states in the computations, (iii) including `hairpin' quark lines in the computations, (iv) developing methods to extract many excited states, and (iv) developing formalism to deal with complex multichannel systems.

These are formidable problems, but steady progress is being made. Recent substantive progress in areas (iii) and (iv) are displayed in a computation of the isoscalar meson spectrum\cite{jlab}.
Similarly, the hadronic scattering problem is being addressed. For example, isovector $\pi\pi$ scattering phase shifts have been recently computed yielding $\rho$ resonance parameters of $m_\rho = 863$ MeV, $g_{\rho\pi\pi} = 4.83$, and $\Gamma_\rho = 10$ MeV. These results have been obtained with a $391$ MeV pion\cite{latt-pipi}. 

Finally, the $D_{s0}$ meson represents a serious challenge for lattice computations since it lies near $DK$ threshold, so long distance effects are important, yet short distance gluon-mediated mass-dependence is also expected to affect the structure of 
the $D_{s0}$\cite{ls}. As a result, old lattice computations of the mass of this state tended to agree with simple quark model estimates (and hence were approximately 200 MeV high). However a new lattice calculation is able to obtain a light $D_{s0}$ state by including the $DK$ continuum in the interpolator set; they also work in $N_f=2+1$ QCD and have a light pion with mass $156$ MeV\cite{mohler}.

\section{Unresolved Puzzles, Old and New}

A rather long list of problems in hadronic physics have resisted progress. Sometimes this is due to lack of experimental facilities and sometimes it is due to our immature theoretical toolkit. At the simplest level many of the new states are of dubious reliability and await confirmation or further exploration. Among these I include 
$G(3900)$, $Y(4008)$, $X(4160)$, $Y(4274)$, $X(4350)$, $Y(4320)$, and $X(4630)$.

The old successes of NRQCD in explaining prompt $J/\psi$ production\cite{lc} have been extended to prompt $\chi{c1}$ and $\chi_{c2}$ production by ATLAS\cite{price}. However NRQCD predictions for $J/\psi$ polarisation have not been successful\cite{pol}. This problem persists at the LHC\cite{lhcb4}. Alternatively, $\Upsilon(1S)$ and $\Upsilon(2S)$ polarisations measured at CMS agree reasonably well with theory, while, curiously, $\Upsilon(3S)$ polarisation fails\cite{uppol}. There is also an interesting attempt by ATLAS to measure prompt production of $J/\psi +W$ which yields results dramatically different from both colour singlet and colour octet models\cite{price}. Finally, CMS have measured the ratio $\sigma(\chi_{b2})\, B(\chi_{b2} \to \Upsilon \gamma)/ \sigma(\chi_{b1})\, B(\chi_{b1} \to \Upsilon \gamma)$ and find quite different behaviour with respect to transverse $\Upsilon$ momentum than expected (see the first of Ref. \cite{uppol}). One can of course add the old problem of NRQCD computations of the rate for $e^+e^- \to J/\psi H$ at the $B$ factories that fall an order of magnitude short.

In addition to the issues with some NRQCD computations, a long series of oddities has persisted in the field. Among these I list

\begin{itemize}
\item the $ee$ widths of $\psi(2S)$ and $\psi(3770)$, which disagree with simple quark model expectations.

\item the ratio of branching ratios\cite{br}

$$\frac{B(J/\psi \to \gamma\eta)}{B(J/\psi \to \gamma \eta')} = 0.21(4)$$
while
$$\frac{B(\psi(2S) \to \gamma\eta)}{B(\psi(2S) \to \gamma \eta')} < 0.018.$$

\item the similar ratios\cite{rm}

$$\frac{B(J/\psi \to \omega\eta)}{B(J/\psi \to \omega \eta')} = 9.56(16)$$
while
$$\frac{B(\psi(2S) \to \omega\eta)}{B(\psi(2S) \to \omega \eta')} < 0.343.$$

\item the ratio of $B$ decays\cite{fb}.

$B\to \eta K \gg B\to \eta'K $ while $ B \to \eta K^* \ll B \to \eta' K^*$.

\item the $\pi-\rho$ puzzle\cite{mo}.

\item node filtering in $J/\psi$ decays. The Dalitz plot for $J/\psi\to \pi\pi\pi$ exhibits a strong $\rho$ signal, while the Dalitz plot for $\psi(2S) \to \pi\pi\pi$ shows a suppressed $\rho$ and a strong $\rho'$. The same happens for the $K\overline{K}\pi$ final state (with $K^*$ and $K^{*'}$ mesons taking the place of the $\rho$ and $\rho'$)\cite{rm}.

\item spin flip in the $\Upsilon$ decay. 
$$
\frac{\Gamma(\Upsilon(5S)\to h_b(nP)\pi\pi)}{\Gamma(\Upsilon(5S)\to \Upsilon(2S)\pi\pi)} = \left\{\begin{array}{ll}0.407(79)(60) & h_b(1P) \\ 0.78(9)(15) & h_b(2P)\end{array}\right.
$$
The numerator involves a $b$-quark spin flip and thus this ratio should be strongly suppressed, yet both are of order unity.

\item oddities in $\Upsilon$ decays: the branching fraction, $B(\Upsilon(5S) \to B^*\overline{B}\pi) = 7.3(2.3)(0.8)$ is an order of magnitude larger than expected\cite{up5s}.

\item oddities in $\Upsilon$ decays: the rates for $\Upsilon(5S)\to \Upsilon(1S, 2S, 3S)\pi\pi$ are two orders of magnitude larger than those of $\Upsilon(2S, 3S, 4S) \to \Upsilon(1S)\pi\pi$\cite{mizuk}.

\item the pQCD prediction 

$$\frac{\Gamma(\chi_{c2}\to \gamma\gamma)}{\Gamma(\chi_{c0}\to \gamma\gamma)}  = \frac{4}{15}(1-1.76\,\alpha_s)$$
disagrees with the measurement of 0.278(50)(18)(31) (note that 4/15 = 0.27)\cite{chis}.

\item the branching fraction $J/\psi \to \gamma\gamma\gamma$ has been measured by CLEO\cite{3g} and indicates large NLO corrections to pQCD computations\cite{3g}.

\end{itemize}

\section{Conclusions}

It is clear that hadronic spectroscopy remains a source of much information -- and confusion -- on the QCD sector of the Standard Model. A survey of the issues raised here indicate that 

\begin{itemize}
\item pQCD appears to largely fail in computations of exclusive processes,
\item the $\pi-\rho$ puzzle and the decays of the $J/\psi$ and $\psi(2S)$ indicate that we do not completely understand the process of hadronisation or factorisation,
\item NRQCD appears to fail in several areas, indicating perhaps that the charm quark is not sufficiently heavy to provide an accurate expansion in NRQCD,
\item bound states continue to confound model builders. Do four-quark states exist? If so, can we model them robustly, can we obtain them on the lattice?
\end{itemize}

On the other side of the ledger, lattice theorists continue to improve their craft and computers continue to get faster; effective field theory has evolved into a mature subject; and many ideas are being generated by the community. It is certain that QCD will continue to surprise and delight us for many years to come.

\noindent
{\bf Acknowledgements}

This research is supported by the US DOE under grant DE-FG02-00ER41135. I am grateful to
Veljko Dmitra\v{s}inovi\'{c}, Ryan Mitchell, Shoichi Sasaki, and Qiang Zhao for informative discussions.

\end{document}